\documentclass[twocolumn,aps,prl,final,superscriptaddress,showpacs]{revtex4-1}
\usepackage[latin9]{inputenc}
\usepackage[active]{srcltx}
\usepackage{color}
\usepackage{amsmath}
\usepackage{graphicx}
\usepackage[unicode=true,pdfusetitle,
 bookmarks=true,bookmarksnumbered=true,bookmarksopen=true,bookmarksopenlevel=1,
 breaklinks=true,pdfborder={0 0 1},backref=false,colorlinks=true]
 {hyperref}
\hypersetup{
 linkcolor=blue,urlcolor=blue,citecolor=blue,pdfstartview={FitH},hyperfootnotes=false}

\makeatletter
\usepackage{dcolumn}% Align table columns on decimal point
\usepackage{bm}% bold math
\usepackage{times}

\makeatother

\begin{document}
\title{Energy storage properties of ferroelectric nanocomposites}
\author{Zhijun Jiang}
\email{zjjiang@xjtu.edu.cn}

\affiliation{MOE Key Laboratory for Nonequilibrium Synthesis and Modulation of
Condensed Matter, School of Physics, Xi'an Jiaotong University, Xi'an
710049, China}
\author{Zhenlong Zhang}
\affiliation{MOE Key Laboratory for Nonequilibrium Synthesis and Modulation of
Condensed Matter, School of Physics, Xi'an Jiaotong University, Xi'an
710049, China}
\author{Sergei Prokhorenko}
\affiliation{Physics Department and Institute for Nanoscience and Engineering,
University of Arkansas, Fayetteville, Arkansas 72701, USA}
\author{Yousra Nahas}
\affiliation{Physics Department and Institute for Nanoscience and Engineering,
University of Arkansas, Fayetteville, Arkansas 72701, USA}
\author{Sergey Prosandeev}
\affiliation{Physics Department and Institute for Nanoscience and Engineering,
University of Arkansas, Fayetteville, Arkansas 72701, USA}
\author{Laurent Bellaiche}
\email{laurent@uark.edu}

\affiliation{Physics Department and Institute for Nanoscience and Engineering,
University of Arkansas, Fayetteville, Arkansas 72701, USA}
\begin{abstract}
An atomistic effective Hamiltonian technique is used to investigate
the finite-temperature energy storage properties of a ferroelectric
nanocomposite consisting of an array of BaTiO$_{3}$ nanowires embedded
in a SrTiO$_{3}$ matrix, for electric field applied along the long
axis of the nanowires. We find that the energy density \textit{versus}
temperature curve adopts a nonlinear, mostly temperature-independent
response when the system exhibits phases possessing an out-of-plane
polarization and vortices while the energy density more linearly increases
with temperature when the nanocomposite either only possesses vortices
(and thus no spontaneous polarization) or is in a paraelectric and
paratoroidic phase for its equilibrium state. Ultrahigh energy density
up to $\simeq$140 J/cm$^{3}$ and an ideal 100\% efficiency are also
predicted in this nanocomposite. A phenomenological model, involving
a coupling between polarization and toroidal moment, is further proposed
to interpret these energy density results. 
\end{abstract}
\maketitle

\section{Introduction}

Dielectric capacitors with high energy densities and efficiencies
are particularly promising for advanced electronics and electric power
systems due to their ultrafast charging/discharging rates \cite{Chu2006,Li2015,Prateek2016,Palneedi2018,Yang2019}.
However, traditional commercial dielectric capacitors, such as biaxially
oriented polypropylene (BOPP), possess relatively low energy density
of about 1.2 J/cm$^{3}$ \cite{Rabuffi2002} while intensive works
have been devoted to improve their energy densities and efficiencies.
One key parameter for energy storage is the recoverable energy density,
which is defined as $U=\int_{P_{\textrm{r}}}^{P_{\textrm{max}}}{\cal E}dP$
\cite{Palneedi2018}, where $P_{\textrm{max}}$ is the maximum polarization
at the maximal applied field, ${\cal E}_{\textrm{max}}$, and $P_{\textrm{r}}$
is the remnant polarization under zero electric field. Another key
parameter is the efficiency $\eta$, defined as $\eta=[U/(U+U_{\textrm{loss}})]\times100\%$
\cite{Palneedi2018}, where $U_{\textrm{loss}}$ is the dissipated
energy because of hysteresis loss and is associated with the area
inside the polarization-\textit{versus}-electric field ($P$-${\cal E}$)
hysteresis loop.

In the last decade, ferroelectric thin films, dielectrics, antiferroelectrics,
relaxor ferroelectrics, superlattices, and lead-free paraelectrics
have been intensively studied in the search of large energy densities
and efficiencies \cite{Pan2019,Peng2015,Xu2017,Xu2022,Instan2017,Kim2020,Pan2021,Jiang2022,Aramberri2022,Jiang2021,Hou2017}.
For instance, a ultrahigh energy density of 112 J/cm$^{3}$ with a
high energy efficiency of 80\% has been observed in lead-free ferroelectric
BiFeO$_{3}$-BaTiO$_{3}$-SrTiO$_{3}$ films \cite{Pan2019}. For
antiferroelectrics, a giant energy density of 154 J/cm$^{3}$ and
97\% efficiency has been achieved in epitaxial lead-free thin films
\cite{Peng2015}. Moreover, relaxor ferroelectrics can also possess
ultrahigh energy densities up to 156 J/cm$^{3}$ and efficiencies
above 90\% \cite{Instan2017,Kim2020,Pan2021,Jiang2022}. Epitaxial
and initially nonpolar AlN/ScN superlattices have also been predicted
to have ultrahigh energy density up to 200 J/cm$^{3}$ with an ideal
efficiency of 100\% \cite{Jiang2021}.

Furthermore, ferroelectric nanocomposites combining ceramic filler
and polymer matrix have shown great potential for high energy storage
capacitors because of their high breakdown strength and high dielectric
permittivity \cite{Li2009,Tang2013,Huang2015}. Experimentally, nanocomposites
made of Ba$_{0.2}$Sr$_{0.8}$TiO$_{3}$ nanowires were shown to reach
a high energy density of 14.86 J/cm$^{3}$ at $4.5\times10^{8}$ V/m.
Based on phase field calculations, Liu \textit{et al}. also numerically
found an energy density of 5 J/cm$^{3}$ and over 95\% high energy
efficiency at a relatively low electric field of 140 MV/m, in nanocomposites
consisting of ferroelectric BaTiO$_{3}$ filler embedded in a polymer
matrix \cite{Liu2017}.

Interestingly, using an atomistic effective Hamiltonian simulations,
different phases were predicted in ferroelectric nanocomposites consisting
of periodic arrays of BaTiO$_{3}$ nanowires embedded in a SrTiO$_{3}$
matrix, for different temperature regions \cite{Louis2012}. Some
of these phases have a coupled macroscopic polarization and an electrical
toroidal moment associated with vortices at low and intermediate temperatures
\cite{Prosandeev2013} while heating the system leads to the progressive
disappearance of the polarization and then vortices (note that frustration
and ordering of topological defects were also found there \cite{Yousra2016}).
One may therefore wonder how these phases, as well as the coupling
between polarization and electrical toroidal moment, affect energy
storage properties in ferroelectric nanocomposites. Also, can these
properties be large? Is it also possible develop a simple model to
analyze and explain their energy density, which may help in designing
future ferroelectric nanocomposite systems with large energy storage
performance?

The aim of this work is to address all the aforementioned important
issues by conducting atomistic first-principles-based effective Hamiltonian
simulations and interpreting the energy storage results via a phenomenological
model. Such simulations and phenomenological model allow us to obtain
a deep insight into energy storage properties of nanostructures. In
particular, ultrahigh energy densities (up to 141.2 J/cm$^{3}$) with
an ideal efficiency of 100\% is presently found. We also demonstrate
that the energy density of ferroelectric nanocomposites can be decomposed
into three energy contributions, each associated with a different
behavior as a function of temperature for different equilibrium phases.
This article is organized as follows. Section II describes details
about the effective Hamiltonian scheme used here. Results are presented
in Sec. III. Finally, a summary is provided in Sec. IV.

\section{Methods}

Here, we use the first-principles-based effective Hamiltonian ($H_{\textrm{eff}}$)
approach developed in Ref. \cite{Walizer2006}, with the total internal
energy $E_{\textrm{int}}$ being written as a sum of two main terms:

\begin{equation}
E_{\textrm{int}}=E_{\textrm{ave}}(\{\mathrm{\mathbf{u}}_{i}\},\{\eta_{I}\},\{\eta_{H}\})+E_{\textrm{\ensuremath{\textrm{loc}}}}(\{\mathrm{\mathbf{u}}_{i}\},\{\eta_{I}\},\{\sigma_{j}\},\{\eta_{\textrm{loc}}\}),\label{eq:E_tot}
\end{equation}
where the first energy term $E_{\textrm{ave}}$ is associated with
the local soft mode $\{\mathrm{\mathbf{u}}_{i}\}$ in unit cell $i$
(that is directly proportional to the electric dipole moment centered
on Ti site $i$), and on the $\{\eta_{I}\}$ and $\{\eta_{H}\}$ inhomogeneous
and homogeneous strain tensors, respectively. $E_{\textrm{ave}}$
consists of five energetic parts: (i) a local mode self-energy; (ii)
the long-range dipole-dipole interaction; (iii) short-range interactions
between local soft modes; (iv) an elastic energy; and (v) interactions
between local modes and strains \cite{Zhong1995}. The second energy
term, $E_{\textrm{loc}}$, involves the $\{\sigma_{j}\}$ and $\{\eta_{\textrm{loc}}\}$
parameters. $\{\sigma_{j}\}$ characterizes the atomic configuration
of the $A$ sublattice, that is $\sigma_{j}$ $=$ $+$1 or $-$1
corresponds to the distribution of Ba or Sr ions located at the $j$
sites of the $A$ sublattice in (Ba$_{x}$Sr$_{1-x}$)TiO$_{3}$ systems,
respectively. $\{\eta_{\textrm{loc}}\}$ represents the local strain
stemming from the difference in ionic radii between Ba and Sr atoms
(that is relatively large $\simeq$2\%). We presently employ this
effective Hamiltonian scheme within Monte Carlo (MC) simulations and
large supercells to obtain energy storage properties in a ferroelectric
BaTiO$_{3}$-SrTiO$_{3}$ nanocomposite. Note that $E_{\textrm{int}}$
of Eq.~(\ref{eq:E_tot}) is used in Monte Carlo simulations with
the Metropolis algorithm \cite{Metropolis1953}, which allows to compute
finite-temperature properties of ferroelectric nanocomposites. Note
also that $E_{\textrm{\ensuremath{\textrm{loc}}}}$ of Eq.~(\ref{eq:E_tot})
automatically implied that intrinsic effects of the interface on physical
properties (such as local electric dipoles and local strains) are
accounted for. On the other hand, the role of structural defects such
as dislocations are not included.

\begin{figure}
\includegraphics[width=8cm]{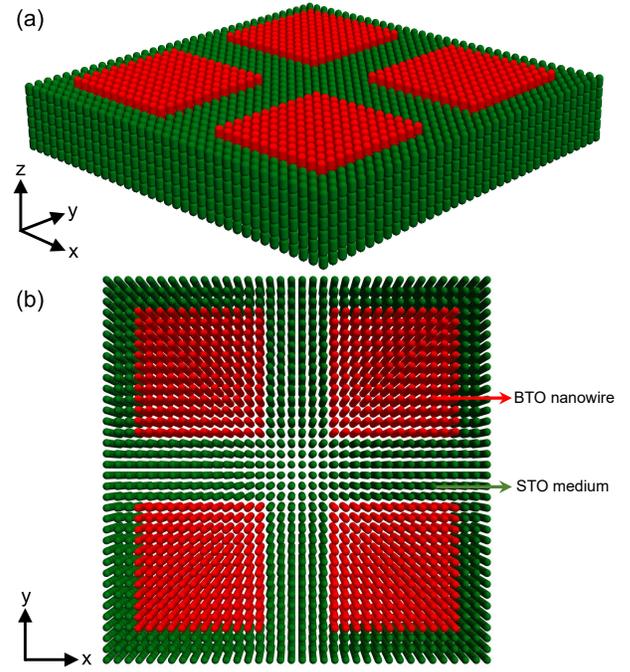}

\caption{Schematic representation of the 36 $\times$ 36 $\times$ 6 supercell
mimicking the studied nanocomposite. (a) The structure is comprised
of four BaTiO$_{3}$ nanowires (red color) with each one having a
cross-sectional of 12 $\times$ 12 (144 Ti sites) along the $x$-
and $y$-directions separated by six sites of SrTiO$_{3}$ medium
(green tubes), with a periodicity of six Ti sites along the $z$-axis
({[}001{]} pseudocubic direction). (b) The top view of the ferroelectric
nanocomposite supercell. \label{fig:structure}}
\end{figure}

Practically, we consider a ferroelectric nanocomposite system made
of a periodic square array of BaTiO$_{3}$ (BTO) nanowires embedded
in a SrTiO$_{3}$ (STO) medium \cite{Louis2012}. Figure~\ref{fig:structure}
shows the considered nanocomposite structure used in this study. Note
that each wire of this nanocomposite has a 4.8 $\times$ 4.8 nm$^{2}$
(144 sites of BTO) rectangular ($x$, $y$) cross section and a long
axis running along the $z$ axis ($x$, $y$, and $z$ axes are parallel
to the pseudocubic {[}100{]}, {[}010{]}, {[}001{]} directions, respectively).
Adjacent wires are separated by 6 sites ($\simeq$2.4 nm) of SrTiO$_{3}$
medium. This nanocomposite is mimicked by a 36 $\times$ 36 $\times$
6 supercell (that contain 38,880 atoms), with a periodicity of 6 sites
($\simeq$2.4 nm) along the $z$ axis.

\begin{figure}
\includegraphics[width=8cm]{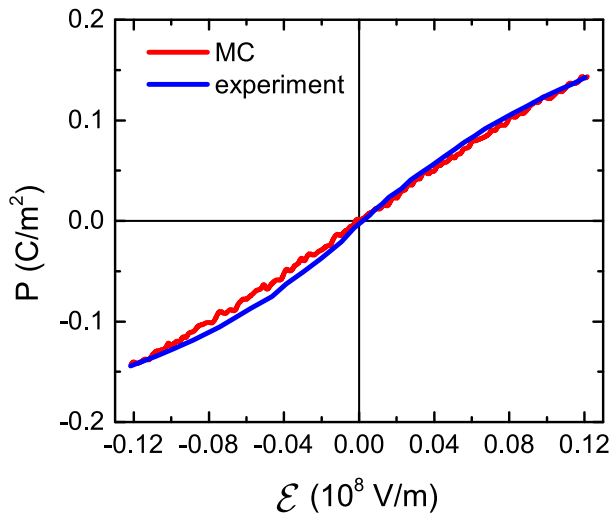}

\caption{$P$-${\cal E}$ hysteresis loops obtained from MC data and from measurements
in (Ba$_{0.5}$Sr$_{0.5}$)TiO$_{3}$ system at 300 K (note that the
theoretical electric field has been divided by a factor of 1.3). \label{fig:P-E_renormalize}}
\end{figure}

To mimic the energy storage properties under an applied dc electric
field, an additional term $-\sum_{i}{\bf p}_{i}\cdot{\bf {\cal E}}$
is added to the total internal energy $E_{\textrm{int}}$, where ${\bf p}_{i}$
is the local electric dipole (which is equal to the product between
the local soft mode $\mathbf{u}_{i}$ and its Born effective charge
$Z^{*}$), and ${\bf {\cal E}}$ is the electric field that is applied
along the $z$ axis. To obtain converged results, 20,000 MC sweeps
are run for equilibration and an additional 20,000 MC sweeps are used
to get the statistical thermal averages at each considered temperature
and applied electric field. Note that we numerically found that the
theoretical electric field is larger from the measured one by a factor
of 1.3 in (Ba$_{x}$Sr$_{1-x}$)TiO$_{3}$ compounds, by comparing
the $H_{\textrm{eff}}$-obtained $P$-${\cal E}$ loop with the experimental
one for disordered (Ba$_{0.5}$Sr$_{0.5}$)TiO$_{3}$ solid solutions
at 300 K \cite{Lu2018}. To correct for such discrepancy, the electric
fields considered in the present study are divided by a factor of
1.3. Figure~\ref{fig:P-E_renormalize} shows the resulting renormalized
$P$-${\cal E}$ loop of the disordered (Ba$_{0.5}$Sr$_{0.5}$)TiO$_{3}$
system at room temperature, which matches the experimental one rather
well. Note that such rescaling is an approach that has been previously
successful in several compounds \cite{Xu2017,Jiang2022,Jiang2021,Jiang2018}. 

\section{Results and discussion}

\subsection{Different phases in the chosen BaTiO$_{3}$-SrTiO$_{3}$ nanocomposite}

\begin{figure}
\includegraphics[width=8cm]{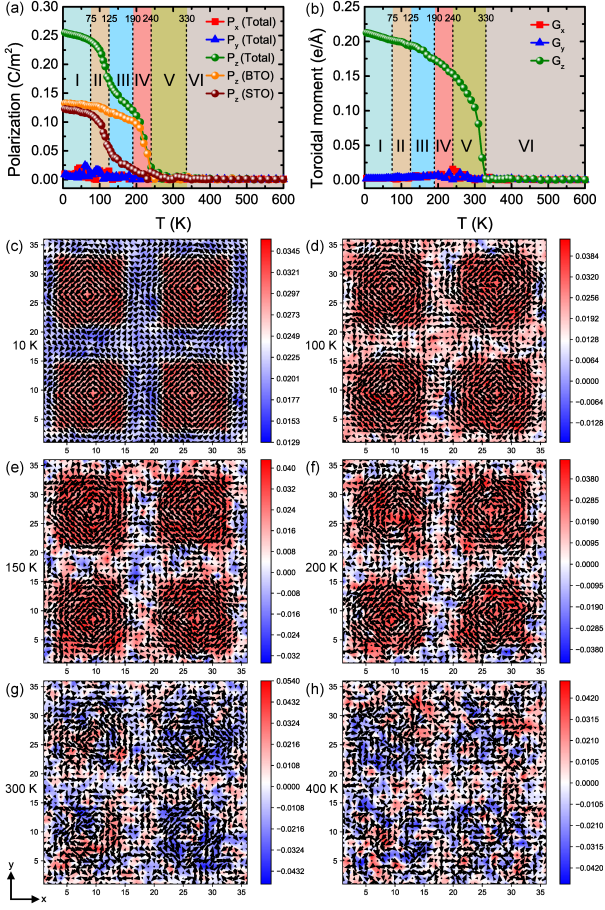}

\caption{Temperature dependence of some properties in the studied BaTiO$_{3}$-SrTiO$_{3}$
nanocomposite: (a) the macroscopic polarization, as well as the contribution
of the $z$-component of the polarization in the BaTiO$_{3}$ wires
and SrTiO$_{3}$ medium to the overall polarization; (b) the average
electrical toroidal moment in the BaTiO$_{3}$ nanowires; and (c)-(f)
snapshots of dipolar configurations in a given ($x$, $y$) plane
at 10 K (Phase I), 100 K (Phase II), 150 K (Phase III), 200 K (Phase
IV), 300 K (Phase V), and 400 K (Phase VI) under zero electric field,
respectively. The color bars indicate the magnitude of the out-of-plane
component of the local modes. \label{fig:finite-temperature-properties}}
\end{figure}

Figures~\ref{fig:finite-temperature-properties}(a)-\ref{fig:finite-temperature-properties}(h)
show the temperature dependence of the overall and individual polarizations,
the electrical toroidal moment, and the dipolar configurations in
a given ($x$, $y$) plane for different temperatures in the chosen
BaTiO$_{3}$-SrTiO$_{3}$ nanocomposite. The polarization contributions
of BTO wires and STO medium to the total $z$-component polarization
in Fig.~\ref{fig:finite-temperature-properties}(a) are given by
$P_{z}\textrm{(BTO)}=a_{\textrm{lat}}Z^{*}\sum{\bf u}_{\textrm{BTO}}/NV$
and $P_{z}\textrm{(STO)}=a_{\textrm{lat}}Z^{*}\sum{\bf u}_{\textrm{STO}}/NV$,
where $a_{\textrm{lat}}$ is the five-atom lattice constant, $Z^{*}$
represents the Born effective charge associated with the local mode,
$N$ is the number of sites in the supercell, $V$ is the unit cell
volume, and $\sum{\bf u}_{\textrm{BTO}}$ and $\sum{\bf u}_{\textrm{STO}}$
are the sum of the local modes centered on BTO wires and STO medium,
respectively. Note that the electrical toroidal moment is defined
as $\mathbf{G}_{j}=\frac{1}{2N_{j}}\sum_{i,j}\mathbf{r}_{i,j}\times\mathbf{p}_{i,j}$,
where $N_{j}$ is the number of sites in nanowire $j$; $\mathbf{p}_{i,j}$
is the local electrical dipole of site $i$ in wire $j$, which is
located at $\mathbf{r}_{i,j}$. A nonzero value of $\mathbf{G}_{j}$
typically characterizes dipole vortex in the nanowire $j$ \cite{Naumov2004},
and the data of Fig.~\ref{fig:finite-temperature-properties}(b)
represents the average of these $\mathbf{G}_{j}$ over the four BaTiO$_{3}$
wires. Based on the evolutions of polarizations and toroidal moment
\textit{versus} temperature, six different phases are identified for
this nanocomposite system. For instance, Phase I exhibits a significant
polarization and toroidal moment in BTO wires both along the pseudocubic
{[}001{]} direction while the STO medium possesses vortices and antivortices
in addition to a polarization along {[}001{]} {[}see Fig.~\ref{fig:finite-temperature-properties}(c){]};
Phase II still has both vortices and polarization along the {[}001{]}
direction in the BTO nanowire, and antivortices and polarization still
occur in the STO medium. However, the $z$-component of the dipoles
in the STO medium is significantly reduced in Phase II {[}see Figs.~\ref{fig:finite-temperature-properties}(a)
and \ref{fig:finite-temperature-properties}(d){]}; In Phase III,
the polarization and vortices still appear in the BTO nanowires, but
the vortices and antivortices in the STO medium have basically disappeared
{[}see Fig.~\ref{fig:finite-temperature-properties}(e){]}; Phase
IV distinguishes itself from Phase III by the annihilation of the
$z$-component of the electrical dipoles in the STO medium {[}see
Figs.~\ref{fig:finite-temperature-properties}(a) and \ref{fig:finite-temperature-properties}(f){]}.
In Phase V, the overall polarization vanishes, which indicates that
the polarization disappears in both BTO wires and STO medium {[}see
Fig.~\ref{fig:finite-temperature-properties}(a){]}. However, the
vortices still exist in the BTO nanowires in Phase V {[}see Fig.~\ref{fig:finite-temperature-properties}(g){]}
as consistent with the nonzero $z$-component of the electrical toroidal
moment {[}see Fig.~\ref{fig:finite-temperature-properties}(b){]};
The paraelectric and paratoroidic Phase VI occurs above 330 K where
both the overall polarization and electrical toroidal moment have
vanished {[}see Figs.~\ref{fig:finite-temperature-properties}(a),
\ref{fig:finite-temperature-properties}(b), and \ref{fig:finite-temperature-properties}(h){]}.
Note that our predicted phases and their temperature range shown in
Fig.~\ref{fig:finite-temperature-properties} are in rather good
agreement with previous theoretical findings \cite{Louis2012} except
for adding the (previously overlooked) new Phase IV where the polarization
in the STO medium has disappeared. Note also that the temperatures
at which successive changes in phases happen are 75, 125, 190, 240,
and 330 K from Phase I to Phase VI, respectively, as emphasized in
Figs.~\ref{fig:finite-temperature-properties}(a) and \ref{fig:finite-temperature-properties}(b).
In order to determine the boundaries between Phases I-IV, we identified
the temperatures at which the in-plane and out-of-plane dielectric
responses peak below 240 K---as similar Ref. \cite{Louis2012}. We
also looked at the temperature dependence of electrical toroidal moment
in the BaTiO$_{3}$ nanowires above 240 K, to determine the boundary
between Phases V and VI.

\subsection{Energy storage properties in the BaTiO$_{3}$-SrTiO$_{3}$ nanocomposite}

\begin{figure}
\includegraphics[width=8cm]{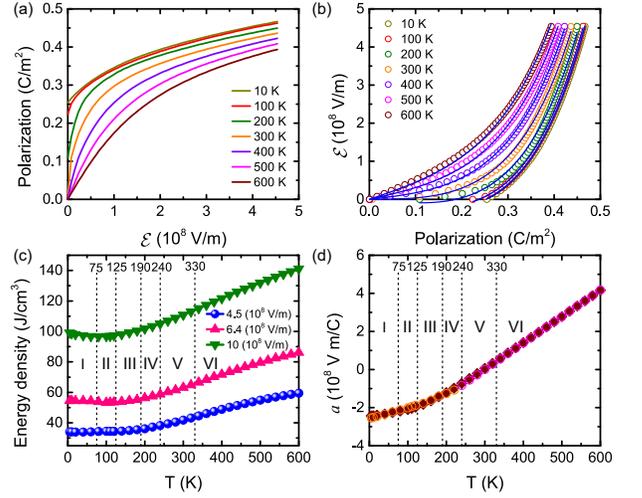}

\caption{(a) $P$-${\cal E}$ curves at different selected temperatures obtained
from MC simulations for electric field applied along the pseudocubic
{[}001{]} direction. (b) Electric field versus polarization at these
different temperatures in BaTiO$_{3}$-SrTiO$_{3}$ nanocomposite.
The solid blue lines represent the fit of the MC ${\cal E}$-$P$
data by Eq.~(\ref{eq:Landau-1}). (c) Energy density obtained from
MC simulations data \textit{versus} temperature for electric fields
applied along the pseudocubic {[}001{]} direction, with maximal applied
electric fields of ${\cal E}_{\textrm{max}}$ $=$ 4.5 $\times$ 10$^{8}$,
6.4 $\times$ 10$^{8}$ and 10 $\times$ 10$^{8}$ V/m, respectively.
(d) Fitting parameter $a$ of Eq.~(\ref{eq:Landau-1}) as a function
of temperature when the maximal applied electric field is equal to
4.5 $\times$ 10$^{8}$ V/m. \label{fig:P-E}}
\end{figure}

In order to investigate the energy storage properties in the BaTiO$_{3}$-SrTiO$_{3}$
nanocomposite, a dc electric field ${\cal E}$ is applied along the
pseudocubic {[}001{]} direction ($z$ axis). Figure~\ref{fig:P-E}(a)
displays the response of the $z$-component of the overall polarization
when the electric field increases from zero to ${\cal E}_{\textrm{max}}$
$=$ $4.5\times10^{8}$ V/m for different temperatures. We also numerically
find that the charging and discharging processes are completely reversible
(note that the charging and discharging correspond to the processes
of increasing the electric field from zero to the maximum applied
field and then decreasing the field back to zero, respectively) for
any considered temperature. The resulting efficiency is therefore
100\%, which has also been reported in epitaxial AlN/ScN superlattices
\cite{Jiang2021} and lead-free Ba(Zr, Ti)O$_{3}$ relaxor ferroelectrics
\cite{Jiang2022} because these two latter compounds possess a field-induced
\textit{second-order} transition from an overall paraelectric to ferroelectric
state. Figure~\ref{fig:P-E}(b) shows the electric field as a function
of polarization for the same temperatures than those indicated in
Fig.~\ref{fig:P-E}(a), which allows us to extract the energy density
$U=\int_{P_{\textrm{r}}}^{P_{\textrm{max}}}{\cal E}dP$ by integrating
the area below the ${\cal E}$-versus-$P$ curve. Such kind of procedure
can be done for any temperature and for any ${\cal E}_{\textrm{max}}$.
Figure~\ref{fig:P-E}(c) displays the resulting energy density as
a function of temperature, when choosing three maximal applied electric
field ${\cal E}_{\textrm{max}}$: $4.5\times10^{8}$, $6.4\times10^{8}$,
and $10\times10^{8}$ V/m. Note that a field of $4.5\times10^{8}$
V/m has been experimentally realized in Ba$_{0.2}$Sr$_{0.8}$TiO$_{3}$
nanocomposites \cite{Tang2013}; $6.4\times10^{8}$ V/m has been reached
for the commercial polypropylene capacitors \cite{Rabuffi2002}; and
a field of $10\times10^{8}$ V/m was reported in La$_{0.1}$Bi$_{0.9}$MnO$_{3}$
and BaTiO$_{3}$ films \cite{Gajek2007,Garcia2009}. As shown in Fig.~\ref{fig:P-E}(c),
the energy density is only slightly temperature-dependent and is nonlinear
below 240 K (these temperatures correspond to the polar Phases I,
II, III and IV). It is equal to 38.3, 58.7, and 105.1 J/cm$^{3}$at
240 K when the maximal applied fields are $4.5\times10^{8}$, $6.4\times10^{8}$,
and $10\times10^{8}$ V/m, respectively. On the other hand, when the
temperature is between 240 and 600 K (which is the range associated
with Phases V and VI), the energy density significantly and more linearly
increases with temperature, which provides values up to 59.5, 86.1
and 141.2 J/cm$^{3}$ for ${\cal E}_{\textrm{max}}$ $=$ $4.5\times10^{8}$,
$6.4\times10^{8}$, and $10\times10^{8}$ V/m, respectively. Strikingly,
the predicted energy densities in the studied BaTiO$_{3}$-SrTiO$_{3}$
nanocomposites therefore exceed the experimentally value of 14.86
J/cm$^{3}$ reported for a maximum electric field of $4.5\times10^{8}$
V/m in Ba$_{0.2}$Sr$_{0.8}$TiO$_{3}$ nanowires \cite{Tang2013},
and is also much larger than the energy density of 1.2 J/cm$^{3}$
achieved in a commercial capacitor with a maximal field of $6.4\times10^{8}$
V/m \cite{Rabuffi2002}.

To understand the energy density behaviors depicted in Fig.~\ref{fig:P-E}(c),
one can use the following simple Landau-type free energy potential:

\begin{equation}
F=\frac{1}{2}a_{0}P^{2}+\frac{1}{4}bP^{4}+\frac{1}{6}cP^{6}+\frac{1}{2}dP^{2}G^{2}-{\cal E}P,\label{eq:Landau}
\end{equation}
where $a_{0}$, $b$, $c$, are coefficients that correspond to quadratic,
quartic, sextic coefficients, respectively, while $d$ quantifies
the sign and strength of the biquadratic coupling between polarization
and electrical toroidal moment.

Under equilibrium condition, the polarization $P$ satisfies $\frac{\partial F}{\partial P}=0$,
which yields:

\begin{equation}
\begin{split}{\cal E} & =(a_{0}+dG^{2})P+bP^{3}+cP^{5}\\
 & =aP+bP^{3}+cP^{5},
\end{split}
\label{eq:Landau-1}
\end{equation}
where $a=a_{0}+dG^{2}$.

As shown in Fig.~\ref{fig:P-E}(b), the electric field \textit{versus}
polarization (${\cal E}$-$P$) data obtained from MC simulations
for all considered temperatures can be relatively well fitted by the
rather simple Eq.~(\ref{eq:Landau-1}) (see solid blue lines), taking
the total polarization from Fig.~\ref{fig:finite-temperature-properties}(a)
and allowing $a$ to be a free parameter while $b$ and $c$ are constant
for any temperature (as consistent with traditional Landau theories).
Such good fitting confirms the validity of our present Landau model,
and also results in the determination of the $b$ and $c$ coefficients
as well as the temperature behavior of $a$. This latter is shown
in Fig.~\ref{fig:P-E}(d) for the maximal field ${\cal E}_{\textrm{max}}$
$=$ $4.5\times10^{8}$ V/m applied along the {[}001{]} pseudocubic
direction. Moreover, this fitting also provides values of the $b$
and $c$ parameters to be 30.9$\times10^{8}$ V m$^{5}$/C$^{3}$
and 114.2$\times10^{8}$ V m$^{9}$/C$^{5}$, respectively.

We will comment on the temperature behavior of the $a$ coefficient
soon, but let us first recall that the recoverable energy density
can be written as $U=\int_{P_{\textrm{r}}}^{P_{\textrm{max}}}{\cal E}dP$,
which, when inserting Eq.~(\ref{eq:Landau-1}), gives:

\begin{equation}
\begin{split}U & =\int_{P_{\textrm{r}}}^{P_{\textrm{max}}}(aP+bP^{3}+cP^{5})dP\\
 & =\frac{1}{2}a(P_{\textrm{max}}^{2}-P_{\textrm{r}}^{2})+\frac{1}{4}b(P_{\textrm{max}}^{4}-P_{\textrm{r}}^{4})+\frac{1}{6}c(P_{\textrm{max}}^{6}-P_{\textrm{r}}^{6}),
\end{split}
\label{eq:energy density Landau}
\end{equation}
where $P_{\textrm{max}}$ is the maximum polarization at ${\cal E}_{\textrm{max}}$
and $P_{\textrm{r}}$ is the remnant polarization. Equation~(\ref{eq:energy density Landau})
therefore tells us that $U$ is a rather straightforward function
of $a$, $b$, $c$, $P_{\textrm{r}}$ and $P_{\textrm{max}}$. Note
that $P_{\textrm{r}}$ is directly obtainable from the MC data, and
is nonzero for temperatures between 5 and 240 K (which covers the
ranges of Phases I, II, III, and IV) while it vanishes for temperatures
above 240 K (corresponding to Phases V and VI) as shown in Figs.~\ref{fig:finite-temperature-properties}(a)
and \ref{fig:P-E}(a). Note also that $P_{\textrm{max}}$ can either
be directly obtained from the MC simulations by taking the value of
the polarization at ${\cal E}_{\textrm{max}}$ or computed via Eq.~(\ref{eq:Landau-1})
at this considered ${\cal E}_{\textrm{max}}$ and using the MC-fitted
parameters of $a$, $b$ and $c$. As we are going to see, both procedures
give similar results.

Let us now comment on the $a$ coefficient. Figure~\ref{fig:P-E}(d)
shows that the fitting parameter $a$ has a nonlinear behavior for
temperatures below 240 K and then basically linearly increases with
temperature between 240 and 600 K, at ${\cal E}_{\textrm{max}}$ $=$
$4.5\times10^{8}$ V/m. For instance, the $a$ parameter decreases
its magnitude from $-$2.44$\times10^{8}$ to $-$0.73$\times10^{8}$
V m/C in a nonlinear fashion between 5 and 240 K, then linearly decreases
in magnitude with temperature between 240 and 600 K up to 4.16$\times10^{8}$
V m/C---with $a$ being equal to zero at 300 K. Note that, as shown
by open circles symbol of Fig.~\ref{fig:P-E}(d), the fitting parameter
$a$ can be well fitted by $a_{1}(T_{c}-T)+dG^{2}$ in the polar Phases
I, II, III and IV below 240 K with $T_{c}$ being equal to 300 K and
the toroidal moment being the one shown in Fig.~\ref{fig:finite-temperature-properties}(b).
The resulting $a_{1}$ is $-4.5\times10^{5}$ V m/C K while $d$ $=$
$183.4$ V $\textrm{Å}^{3}$/$e^{3}$. The positive sign of $d$ therefore
indicates a competition between polarization and toroidal moment,
which also explains why the fitting provides a $T_{c}$ of 300 K,
while the true Curie temperature of the studied nanocomposite is lower
and equal to 240 K. Moreover and as also shown by open circles symbols
in Fig.~\ref{fig:P-E}(d), $a$ can further be well fitted by $a=a_{2}(T-T_{c})$
in Region V (for which the polarization has vanished but the toroidal
moment still exists) and Region VI (for which the total polarization
and toroidal moment have both been annihilated), with $T_{c}$ being
equal to 300 K too. Note that the fitted $a_{2}$ is $1.4\times10^{6}$
V m/C K and that these different behaviors and analytical formula
of $a$ for temperatures below \textit{versus} above 240 K are consistent
with the general line indicated below Eq.~(\ref{eq:Landau-1}), namely
that $a=a_{0}+dG^{2}$---with $a_{0}$ being directly proportional
to $(T-T_{c})$, as consistent with typical Landau theory, and with
$d$ being finite when both spontaneous polarization and toroidal
moment exist and zero otherwise.

\begin{figure}
\includegraphics[width=8cm]{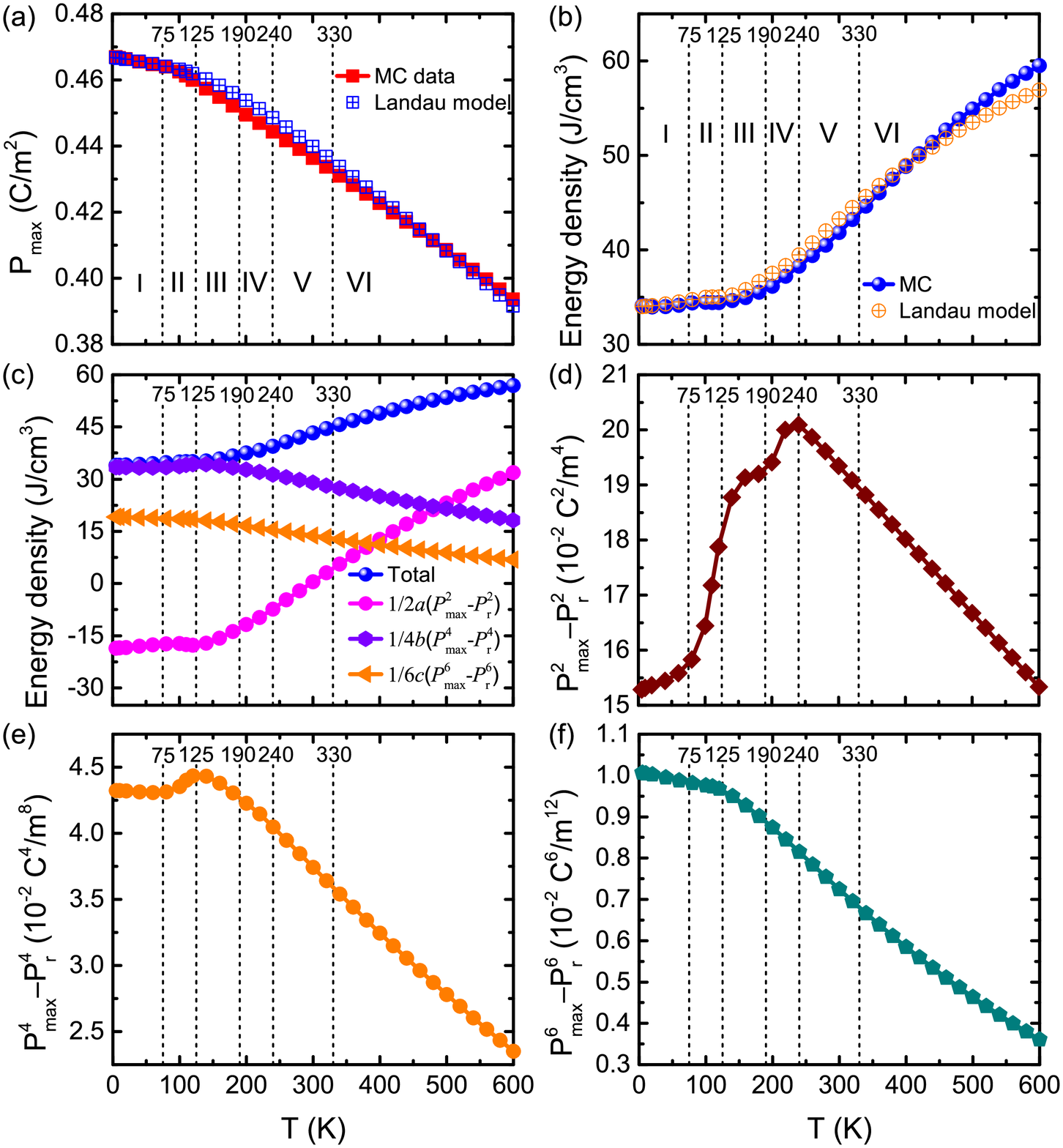}

\caption{(a) The maximum polarization $P_{\textrm{max}}$ obtained from MC
simulations and Landau model {[}see Eq.~(\ref{eq:Landau-1}){]} as
a function of temperature for ${\cal E}_{\textrm{max}}$ $=$ 4.5
$\times$ 10$^{8}$ V/m and fields applied along the {[}001{]} direction
in the studied BaTiO$_{3}$-SrTiO$_{3}$ nanocomposite. (b) Energy
density obtained from MC simulations and Eq.~(\ref{eq:energy density Landau})
as a function of temperature for a maximal applied electric field
${\cal E}_{\textrm{max}}$ $=$ 4.5 $\times$ 10$^{8}$ V/m. (c) The
total and decomposed energy densities $\frac{1}{2}a(P_{\textrm{max}}^{2}-P_{\textrm{r}}^{2})$,
$\frac{1}{4}b(P_{\textrm{max}}^{4}-P_{\textrm{r}}^{4})$, and $\frac{1}{6}c(P_{\textrm{max}}^{6}-P_{\textrm{r}}^{6})$
obtained from Eq.~(\ref{eq:energy density Landau}) as a function
of temperature for ${\cal E}_{\textrm{max}}$ $=$ 4.5 $\times$ 10$^{8}$
V/m. (d)-(f) $P_{\textrm{max}}^{2}-P_{\textrm{r}}^{2}$, $P_{\textrm{max}}^{4}-P_{\textrm{r}}^{4}$,
and $P_{\textrm{max}}^{6}-P_{\textrm{r}}^{6}$ versus temperature
for ${\cal E}_{\textrm{max}}$ $=$ 4.5 $\times$ 10$^{8}$ V/m, respectively.
\label{fig:Pmax}}
\end{figure}

Furthermore, Fig.~\ref{fig:Pmax}(a) displays the value of the maximum
polarization, $P_{\textrm{max}}$ as a function of temperature both
from MC simulations and from the Landau model using the MC-fitted
parameters of $a$, $b$ and $c$ in Eq.~(\ref{eq:Landau-1}). One
can clearly see that for all considered temperatures at ${\cal E}_{\textrm{max}}$
$=$ 4.5 $\times$ 10$^{8}$ V/m, the MC simulations and the Landau-model-obtained
$P_{\textrm{max}}$ provide nearly similar results, which is quite
remarkable once realizing the simplicity of Eq. (3), on one hand,
and the complexity of the investigated system on the other hand. As
shown in Fig.~\ref{fig:Pmax}(a), $P_{\textrm{max}}$ almost linearly
and very slightly decreases with temperature in Regions I and II for
temperatures between 5 and 125 K (values varying between 0.467 to
0.460 C/m$^{2}$). In Regions III, IV, V and VI for temperatures ranging
between 125 and 600 K, $P_{\textrm{max}}$ basically linearly decreases
in a more significant fashion with temperature (the value of $P_{\textrm{max}}$
varies from 0.460 to 0.394 C/m$^{2}$).

To understand the energy density results in Fig.~\ref{fig:P-E}(c),
we take advantage of Eq.~(\ref{eq:energy density Landau}). Figure~\ref{fig:Pmax}(b)
shows the energy density directly obtained from Eq.~(\ref{eq:energy density Landau})
at the maximal applied field of ${\cal E}_{\textrm{max}}$ $=$ 4.5
$\times$ 10$^{8}$ V/m, along with the energy density data computed
from the MC simulations. The Landau-model-obtained energy density
agrees reasonably well with the MC-obtained energy density. Moreover
and according to Eq.~(\ref{eq:energy density Landau}), the energy
density is the sum of three terms, which are $\frac{1}{2}a(P_{\textrm{max}}^{2}-P_{\textrm{r}}^{2})$,
$\frac{1}{4}b(P_{\textrm{max}}^{4}-P_{\textrm{r}}^{4})$, and $\frac{1}{6}c(P_{\textrm{max}}^{6}-P_{\textrm{r}}^{6})$.
The three contributions of energy density are shown in Fig.~\ref{fig:Pmax}(c),
while Figs.~\ref{fig:Pmax}(d), (e) and (f) display the temperature
dependency of $(P_{\textrm{max}}^{2}-P_{\textrm{r}}^{2})$, $(P_{\textrm{max}}^{4}-P_{\textrm{r}}^{4})$
and $(P_{\textrm{max}}^{6}-P_{\textrm{r}}^{6})$, respectively. The
first contribution of the energy density thus relies on the product
of the $a$ parameter and $P_{\textrm{max}}^{2}-P_{\textrm{r}}^{2}$,
and only slightly depends on temperature with a nonlinear behavior
in Regions I and II for temperature between 5 and 125 K (the energy
density value associated with this first term ranges from $-$18.6
to $-$17.7 J/cm$^{\textrm{3}}$ within Phases I and II). This first
energy density term of $\frac{1}{2}a(P_{\textrm{max}}^{2}-P_{\textrm{r}}^{2})$
is almost constant in this temperature range because there is a compensation
between the facts that (the negative) $a$ decreases in magnitude
and that $P_{\textrm{max}}^{2}-P_{\textrm{r}}^{2}$ increases with
temperature. On the other hand, such compensation does not occur anymore
for temperatures from 125 to 240 K (from Phase III to Phase IV) due
to the strong nonlinear increase of $P_{\textrm{max}}^{2}-P_{\textrm{r}}^{2}$
as well as the more pronounced (nonlinear) decrease of the magnitude
of $a$. Consequently, the first energy density term, $\frac{1}{2}a(P_{\textrm{max}}^{2}-P_{\textrm{r}}^{2})$,
almost linearly increases with temperature between 125 and 240 K.
In Phases V and VI (for temperatures between 240 and 600 K), $a$
linearly increases with temperature (from $-$0.73$\times10^{8}$
to 4.16$\times10^{8}$ V m/C) faster than $P_{\textrm{max}}^{2}-P_{\textrm{r}}^{2}$
decreases with temperature (from 0.201 to 0.153 C$^{2}$/m$^{4}$),
hence resulting in $\frac{1}{2}a(P_{\textrm{max}}^{2}-P_{\textrm{r}}^{2})$
increasing in a linear fashion with temperature.

The second contribution of energy density, $\frac{1}{4}b(P_{\textrm{max}}^{4}-P_{\textrm{r}}^{4})$,
is only slightly dependent on temperature between 5 and 125 K (the
values varying between 33.4 and 34.3 J/cm$^{\textrm{3}}$) because
$P_{\textrm{max}}^{4}-P_{\textrm{r}}^{4}$ is basically constant (around
0.043 C$^{4}$/m$^{8}$) there and $b$ is always a constant in our
fitting. Above 125 K (from Region III to Region VI), $\frac{1}{4}b(P_{\textrm{max}}^{4}-P_{\textrm{r}}^{4})$
linearly decreases with temperature up to 600 K (from 34.3 at 125
K to 18.2 J/cm$^{3}$ at 600 K) because $P_{\textrm{max}}^{4}-P_{\textrm{r}}^{4}$
adopts such behavior (it decreases from 0.044 to 0.024 C$^{4}$/m$^{8}$).

The third energy density, $\frac{1}{6}c(P_{\textrm{max}}^{6}-P_{\textrm{r}}^{6})$,
only very slightly decreases with temperature in Phases I and II,
\textit{i.e.} for temperatures ranging between 5 and 125 K, which
arises from the weak decrease of $P_{\textrm{max}}^{6}-P_{\textrm{r}}^{6}$
(from 0.0101 C$^{6}$/m$^{12}$ at 5 K to 0.0097 C$^{6}$/m$^{12}$
at 125 K) in these regions---since the $c$ parameter is constant
too. Furthermore, for temperatures ranging between 125 and 600 K (in
Regions III, IV, V and VI ), $\frac{1}{6}c(P_{\textrm{max}}^{6}-P_{\textrm{r}}^{6})$
rather strongly and continuously decreases with temperature up to
600 K (from 18.4 to 6.9 J/cm$^{3}$) as a result of the significant
decrease of $P_{\textrm{max}}^{6}-P_{\textrm{r}}^{6}$ with temperature
from 0.0097 to 0.0036 C$^{6}$/m$^{12}$.

Figure~\ref{fig:Pmax}(c) further shows that the energy densities
of $\frac{1}{2}a(P_{\textrm{max}}^{2}-P_{\textrm{r}}^{2})$ and $\frac{1}{6}c(P_{\textrm{max}}^{6}-P_{\textrm{r}}^{6})$
nearly cancel each other in Phases I and II for temperatures ranging
between 5 and 125 K, implying that $\frac{1}{4}b(P_{\textrm{max}}^{4}-P_{\textrm{r}}^{4})$
is the dominant contribution there---which also explains why the
total energy density only very slightly depends on temperatures in
these regions. In Phases III and IV, the first energy density $\frac{1}{2}a(P_{\textrm{max}}^{2}-P_{\textrm{r}}^{2})$
increases with temperature while the second and third energy densities
both decrease, which, once again, results in a total energy density
that only weakly depends on temperature. On the other hand, in Phases
V and VI, the total energy density is significantly enhanced with
temperature, and in a nearly linear fashion, following the strong
linear increase of $\frac{1}{2}a(P_{\textrm{max}}^{2}-P_{\textrm{r}}^{2})$
which is counteracted by the smaller linear decrease of the second
and third energy densities. Interestingly, the contributions in percentage
of the total energy density can be temperature-dependent between Regions
V and VI. As a matter of fact, the contributions of $\frac{1}{2}a(P_{\textrm{max}}^{2}-P_{\textrm{r}}^{2})$,
$\frac{1}{4}b(P_{\textrm{max}}^{4}-P_{\textrm{r}}^{4})$ and $\frac{1}{6}c(P_{\textrm{max}}^{6}-P_{\textrm{r}}^{6})$
to the total energy density at 300 K are 0\%, 67\%, and 33\%, respectively
(the zero value of the first energy term arises from the annihilation
of $a$ at 300 K). This is to be compared with the corresponding numbers
of 56\%, 32\%, and 12\%, respectively, at 600 K.

\section{Summary}

In conclusion, based on atomistic effective Hamiltonian scheme combined
with Monte Carlo simulations, we investigated the energy storage properties
in a BaTiO$_{3}$-SrTiO$_{3}$ nanocomposite consisting of BaTiO$_{3}$
nanowires embedded in a SrTiO$_{3}$ matrix. We found that this nanocomposite
system can exhibit large energy densities and an ideal 100\% efficiency
for three considered maximal applied electric fields. It is also found
that the energy density-\textit{versus}-temperature curve is nonlinear
and only weakly dependent on temperature, for temperatures below 240
K (for which the equilibrium phases are polar). On the other hand,
it becomes more linear and strongly temperature-dependent as the temperature
increases from 240 to 600 K, when the system progressively loses its
spontaneous polarization and then its spontaneous toroidal moment.
Such unusual energy storage features are then interpreted via the
development of a simple Landau model that reproduces the Monte Carlo
simulation data and that also involves a coupling between polarization
and toroidal moment. In particular, the energy density consists of
three energy terms, namely $\frac{1}{2}a(P_{\textrm{max}}^{2}-P_{\textrm{r}}^{2})$,
$\frac{1}{4}b(P_{\textrm{max}}^{4}-P_{\textrm{r}}^{4})$, and $\frac{1}{6}c(P_{\textrm{max}}^{6}-P_{\textrm{r}}^{6})$,
that adopt different behaviors in different structural phases. The
proposed phenomenological model may be further put in use to search
for, or analyze results of, other ferroelectric nanostructures with
large energy density. 
\begin{acknowledgments}
This work is supported by the National Natural Science Foundation
of China (Grant No.\ 11804138), Natural Science Basic Research Program
of Shaanxi (Program No.\ 2023-JC-YB-017), Shaanxi Fundamental Science
Research Project for Mathematics and Physics (Grant No.\ 22JSQ013),
``Young Talent Support Plan'' of Xi'an Jiaotong University (Grant
No.\ WL6J004), and the Fundamental Research Funds for the Central
Universities. L.B.\ acknowledges ARO Grant No.\ W911NF-21-1-0113
and the Vannevar Bush Faculty Fellowship Grant No.\ N00014-20-1-2834
from the Department of Defense. The HPC Platform of Xi'an Jiaotong
University is also acknowledged. 
\end{acknowledgments}

\end{document}